\documentclass[showpacs,preprintnumbers,onecolumn]{revtex4}
\usepackage{amssymb}
\usepackage{amsfonts}
\usepackage{amsmath}
\usepackage{graphicx}
\usepackage{dcolumn}
\usepackage{bm}

\begin{document}

\title{Criticality-induced universality in ratchets}
\author{Ricardo Chac\'{o}n}
\affiliation{Departamento de F\'{\i}sica Aplicada, Escuela de Ingenier\'{\i}as
Industriales, Universidad de Extremadura, Apartado Postal 382, E-06071
Badajoz, Spain}
\date{\today}

\begin{abstract}
Conclusive mathematical arguments are presented supporting the ratchet
conjecture [R. Chac\'{o}n, J. Phys. A \textbf{40}, F413 (2007)], i.e., the
existence of a universal force waveform which optimally enhances directed
transport by symmetry breaking. Specifically, such a particular waveform is
shown to be \textit{unique} for both temporal and spatial biharmonic forces,
and general (\textit{non}-perturbative) laws providing the dependence of the
strength of directed transport on the force parameters are deduced for these
forces. The theory explains previous results for a great diversity of
systems subjected to such biharmonic forces and provides a universal
quantitative criterion to optimize \textit{any} application of the ratchet
effect induced by symmetry breaking of temporal and spatial biharmonic
forces.
\end{abstract}

\maketitle

% Force line breaks with \\




\textbf{\ }Symmetry principles play a fundamental role in the laws of nature
by constituting a synthesis of those regularities that are independent of
the specific dynamics. Since the beginning of scientific thinking, most of
these symmetry principles have been deeply associated with certain
\textquotedblleft principles of economy\textquotedblright\ of nature, such
as the principle of least action and other variational principles [1].
Curie's principle [2] is just one example of such a symmetry principle. It
has been applied for instance to the ubiquitous phenomenon of directed
transport by symmetry breaking without any net external force$-$the
so-called ratchet effect [3,4,5]. Directed ratchet transport (DRT) is
nowadays understood as a result of the interplay of nonlinearity, symmetry
breaking, and non-equilibrium fluctuations, where these fluctuations may
include temporal noise [6], spatial disorder [7], and quenched temporal
disorder [8]. The space-time symmetries of some generic equations of motion,
which have to be broken to allow the appearance of DRT, have recently been
proposed [9,10,11] in accordance with Curie's principle. While this symmetry
analysis provides useful necessary conditions on the ac (temporal) forces
and the static (spatial) potential for DRT to appear, no information at all
concerning its strength can be obtained from such a symmetry analysis. This
is not altogether surprising since Curie's principle \textit{per se}, in
contrast to variational principles, lacks a mathematical formulation
connecting quantitatively a measure of the "degree of symmetry" of causes
with that of effects, thus impeding the observation of its possible
connection (if any) with some natural \textquotedblleft principle of
economy\textquotedblright\ [12]. In other words, Curie's principle must be
considered as belonging to the philosophical rather than the scientific
realm [13]. To overcome this difficulty in the context of DRT, there has
recently been proposed a \textit{quantitative} measure of the degree of
symmetry breaking (DSB) on which the strength of DRT must depend [14]. It
should be stressed that this quantitative relationship between cause
(symmetry breaking) and effect (DRT)$-$the so-called \textit{DSB mechanism}
[14]$-$is absolutely absent in the formulation of Curie's principle. Once
one assumes that the breakage of the relevant (for each particular equation
of motion) space-time symmetries can be quantified, and that the strength of
the resulting DRT (hereafter referred to as $\left\langle V\right\rangle $)
is proportional to this degree of breakage, the following questions
naturally arise: Can one find the parameter values of the periodic zero-mean
forces involved that maximally break the relevant symmetries and thus
optimally enhance DRT? Are these values universal, implying therefore the
existence of a unique optimal force waveform?

In this Letter, a positive response to these questions is provided by
completing the mathematical proof of the ratchet conjecture [14], i.e., the
existence of a universal force waveform which optimally enhances DRT. The
biharmonic temporal force $F(t^{\prime })/\epsilon =\eta \cos \left( \omega
t^{\prime }+\varphi _{1}\right) +\left( 1-\eta \right) \cos \left( 2\omega
t^{\prime }+\varphi _{2}\right) $ and the biharmonic spatial potential $%
U(x)/\epsilon =\eta \sin \left( kx^{\prime }+\varphi _{1}\right) +\left(
1-\eta \right) \sin \left( 2kx^{\prime }+\varphi _{2}\right) $, $\left(
\epsilon >0,\eta \in \left[ 0,1\right] \right) $, have been (and still are)
overwhelmingly used as standard models in analytical research on ratchets
[15]. A simple re-scaling of these biharmonic functions yields%
\begin{eqnarray}
f(t) &=&\eta \cos \left( t\right) +\left( 1-\eta \right) \cos \left(
2t+\varphi _{eff}\right) ,  \label{1} \\
g(x) &=&\eta \cos \left( x\right) +2\left( 1-\eta \right) \cos \left(
2x+\varphi _{eff}\right) ,  \label{2}
\end{eqnarray}%
where $t=\omega t^{\prime }+\varphi _{1},x=kx^{\prime }+\varphi
_{1},f(t)=F(t)/\epsilon ,g(x)=-\left( \epsilon k\right) ^{-1}dU(x)/dx$, and $%
\varphi _{eff}=\varphi _{2}-2\varphi _{1}$ is the effective phase. Thus, the
breakage of the three relevant symmetries associated with the forces (1) and
(2) (i.e., the shift symmetry and the two reversal symmetries, see Refs.
[6,9,10,11]) is controlled by only two parameters: $\eta $ (relative
amplitude of the two harmonics) and $\varphi _{eff}$. But, unfortunately,
changing these parameters implies \textit{also} changing the amplitude and
symmetry of the positive and negative parts of the biharmonic forces with
respect to the symmetric cases $\eta =\left\{ 0,1\right\} $ (see Figs.~1(a),
2(a), and 3(a)). Since the strength of any transport (induced by symmetry
breaking or not, i.e., by non-zero-mean forces) depends upon the amplitude
of the driving forces, one concludes that these two effects of changing $%
\eta $ or $\varphi _{eff}$ overlap, so that one will find it difficult to
distinguish the contribution to transport that is \textit{purely} due to
symmetry breaking. A first step towards clarifying this problematic
situation and answering the above questions has been to study the breakage
of the relevant symmetries of the elliptic force%
\begin{equation}
f_{ellip}(t)={\rm sn}\left( Kt/\pi ;m\right) {\rm cn}\left( Kt/\pi
;m\right) ,  \label{3}
\end{equation}%
where $K\equiv K(m)$ is the complete elliptic integral of the first kind
while ${\rm sn}\left( \cdot ;m\right) $ and ${\rm cn}\left( \cdot
;m\right) $ are Jacobian elliptic functions of parameter $m\in \left[ 0,1%
\right] $ [16]. Now, $m$ is the single parameter controlling the breakage of
the relevant symmetries. Unlike functions (1) and (2), function (3) exhibits
the advantageous property that its waveform changes while its amplitude and
image remain constant, $f_{ellip}(t)\in \left[ -1/2,1/2\right] ,\forall t$,
as the shape parameter $m$ varies from $0$ to $1$ (see Fig.~1 in Ref. [14]).
It has been demonstrated [14] that its maximal (with respect to the
integration range) transmitted impulse over a half-period, $I\left[ f\right]
\equiv \left\vert \int\nolimits_{T/2}f\left( t\right) dt\right\vert $, and
its DSB are monotonously decreasing and increasing functions of $m$,
respectively. This result has led to the conclusion that optimal enhancement
of DRT is achieved when maximal effective (i.e., \textit{critical}) symmetry
breaking occurs, which is in turn a consequence of two reshaping-induced
competing effects: the increase of the DSB and the decrease of the maximal
transmitted impulse over a half-period, thus implying the existence of a
\textit{particular} force waveform which optimally enhances DRT. However,
reaching the same conclusion for the biharmonic force (1) is a harder task
due to the aforementioned problematic situation.

\begin{figure}[htb]
  \centering
  \includegraphics[width=6.5cm]{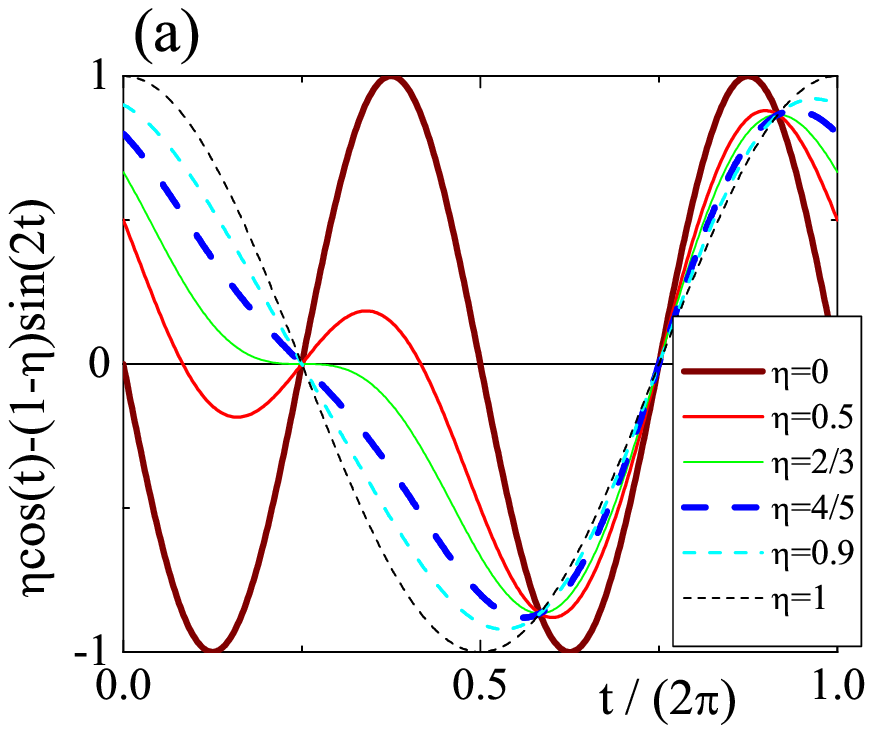}
  \includegraphics[width=6.5cm]{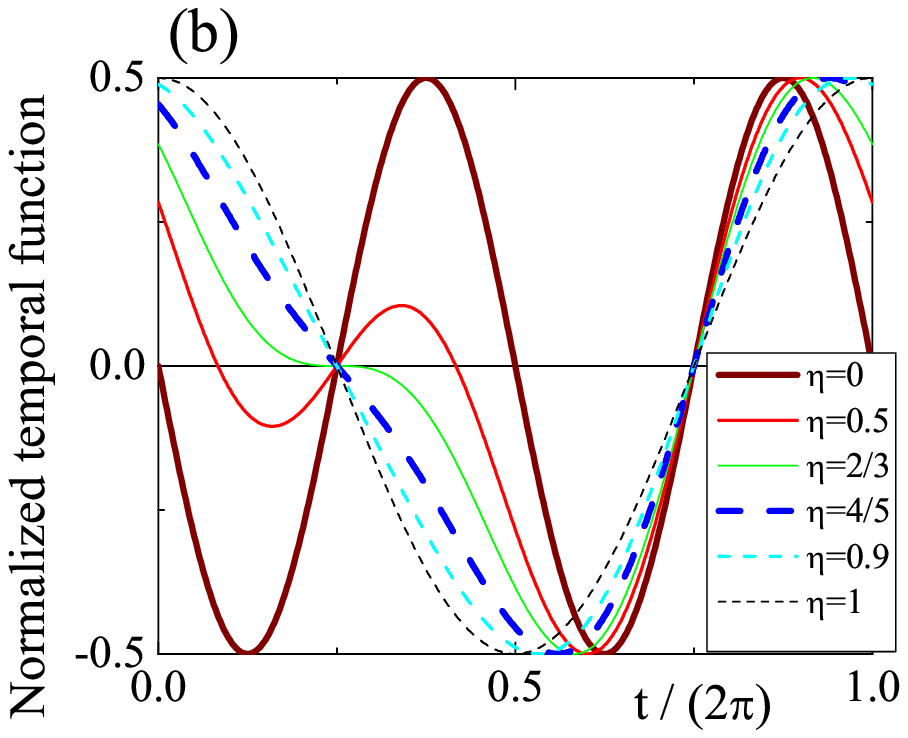}
  \caption{(Color online). (a) Temporal biharmonic function [Eq.~(1)] for the
  optimal value $\varphi _{eff}=\pi /2$, and (b) the corresponding normalized
  function [Eq.~(5)], versus time for different values of $\eta $.
  }
  \label{fig:one}
\end{figure}

\begin{figure}[htb]
  \centering
  \includegraphics[width=6.5cm]{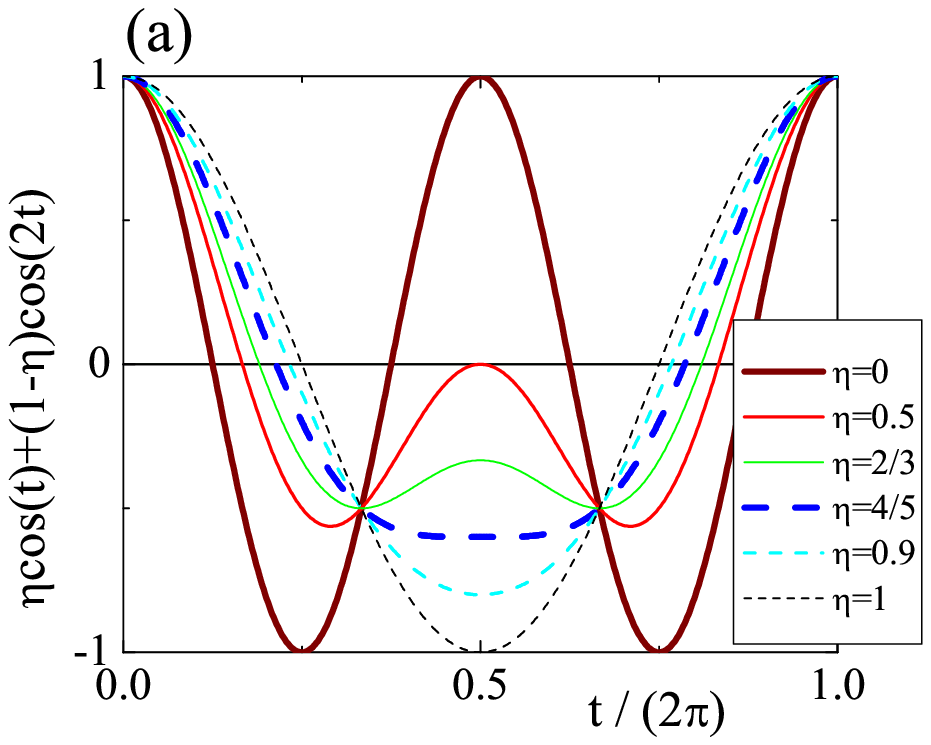}
  \includegraphics[width=6.5cm]{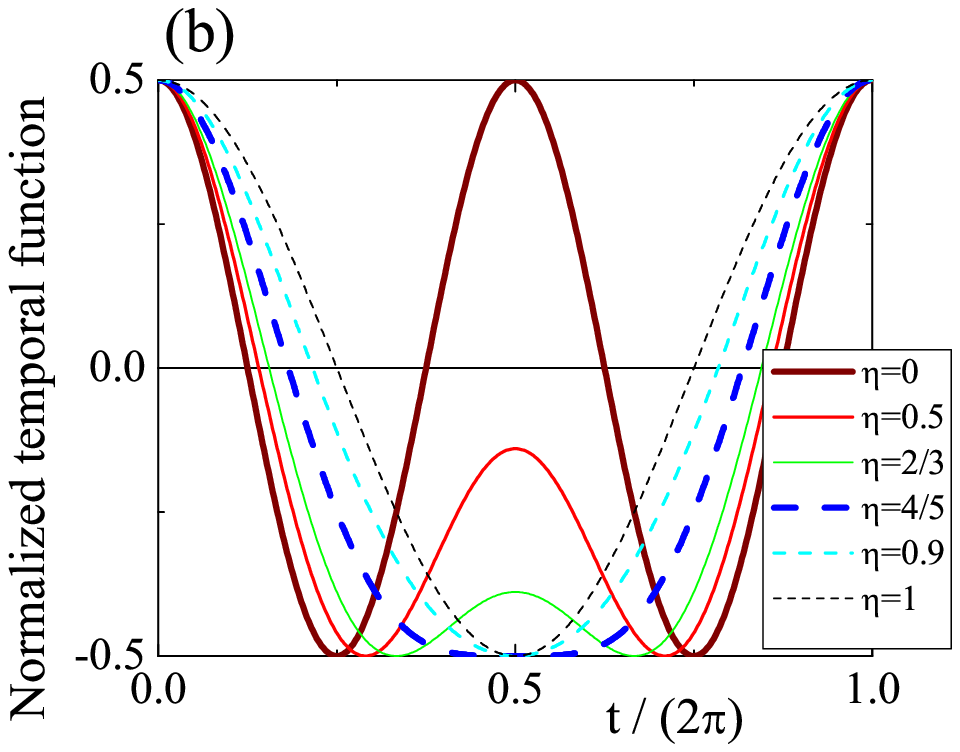}
  \caption{(Color online). (a) Temporal biharmonic function [Eq.~(1)] for the
  least favourable value $\varphi _{eff}=0$, and (b) the corresponding
  normalized function [Eqs.~(7) and (8)], versus time $t$ for different values
  of $\eta $.
  }
  \label{fig:two}
\end{figure}

While it has been shown for Hamiltonian systems subjected to symmetric
spatial potentials and force (1) [14] that when $\left( 1-\eta \right) /\eta
\geqslant 1/2$ and $\varphi _{eff}=\left\{ \pi /2,3\pi /2\right\} $ the DSB
is maximal, it is shown in this Letter that for such optimal values of $%
\varphi _{eff}$ the maximal transmitted impulse over a half-period exhibits
a single maximum at $\left( 1-\eta \right) /\eta =1/2$ (i.e. $\eta =2/3$)
while the amplitude of the force (1) is held constant, concluding again that
optimal enhancement of DRT is achieved when critical symmetry breaking
occurs. Consider, for example, the case $\varphi _{eff}=\pi /2$. Clearly,
one needs an affine transformation to renormalize the function $\eta \cos
\left( t\right) -\left( 1-\eta \right) \sin \left( 2t\right) $ in order to
change its image from $\left[ -M\left( \eta \right) ,M\left( \eta \right) %
\right] $, where%
\begin{eqnarray}
M\left( \eta \right) &\equiv &\frac{3\eta +N(\eta )}{32\left( 1-\eta \right)
}\sqrt{32+30\eta ^{2}+2\eta \left[ N(\eta )-32\right] ,}  \notag \\
&&N(\eta )\equiv \sqrt{32-64\eta +33\eta ^{2}},  \label{4}
\end{eqnarray}%
to $\left[ -1/2,1/2\right] $ for $0\leqslant \eta \leqslant 1$ (in
accordance with both the amplitude and the image of the elliptic force (3)).
One therefore obtains the normalized function%
\begin{equation}
f_{\varphi _{eff}=\pi /2}^{\ast }(t)\equiv \frac{f_{\varphi _{eff}=\pi /2}(t)%
}{2M\left( \eta \right) }.  \label{5}
\end{equation}%
Figure 1(b) shows a plot of $f_{\varphi _{eff}=\pi /2}^{\ast }(t)$. One
readily finds that its corresponding maximal transmitted impulse over a
half-period,%
\begin{equation}
I[f_{\varphi _{eff}=\pi /2}^{\ast }]\left( \eta \right) =\frac{1}{M\left(
\eta \right) },  \label{6}
\end{equation}%
exhibits a single maximum at $\eta =2/3$, confirming thus the above scenario
of optimal enhancement of DRT by critical symmetry breaking (see Fig.~4
below for a plot of $I[f_{\varphi _{eff}=\pi /2}^{\ast }]-1$ versus $\eta $).

Since the breakage of the shift symmetry of any biharmonic function [Eqs.
(1) and (2)] occurs for \textit{all} values of the effective phase, one
could expect on the basis of the DSB mechanism that the above scenario
should hold for any other value of the effective phase, including the least
favourable ones, i.e., those values not breaking the corresponding relevant
reversal symmetry [14]. It is shown here that this is indeed the case by
considering the illustrative value $\varphi _{eff}=0$. Using now a different
affine transformation, one renormalizes the function $\eta \cos \left(
t\right) +\left( 1-\eta \right) \cos \left( 2t\right) $ to change its image
from $\left[ -\frac{9\eta ^{2}-16\eta +8}{8-8\eta },1\right] $ for $%
0\leqslant \eta \leqslant 4/5$ and $\left[ 1-2\eta ,1\right] $ for $%
4/5\leqslant \eta \leqslant 1$ to $\left[ -1/2,1/2\right] $ for $0\leqslant
\eta \leqslant 1$. Noting again that this renormalization process is unique,
one straightforwardly obtains the normalized function%
\begin{eqnarray}
f_{\varphi _{eff}=0}^{\ast }(t) &\equiv &\frac{f_{\varphi
_{eff}=0}(t)-1/2-R\left( \eta \right) /2}{1-R(\eta )},  \label{7} \\
R\left( \eta \right) &\equiv &\left\{
\begin{array}{c}
\frac{9\eta ^{2}-16\eta +8}{8\eta -8},\ 0\leqslant \eta \leqslant 4/5 \\
1-2\eta ,\ 4/5\leqslant \eta \leqslant 1%
\end{array}%
\right\} .  \label{8}
\end{eqnarray}%
A plot of $f_{\varphi _{eff}=0}^{\ast }(t)$ is shown in Fig.~2(b). One finds
that the corresponding maximal transmitted impulse over a half-period,%
\begin{equation}
I[f_{\varphi _{eff}=0}^{\ast }]\left( \eta \right) =\left( \frac{\pi }{2}%
\right) \frac{1+R\left( \eta \right) }{1-R\left( \eta \right) },  \label{9}
\end{equation}%
also exhibits a single maximum at $\eta =2/3$ with $I[f_{\varphi
_{eff}=0}^{\ast }]\left( \eta =2/3\right) =\pi /6$, confirming again the
above scenario of optimal enhancement of DRT by critical symmetry breaking
(see Fig.~4 below for a plot of $I[f_{\varphi _{eff}=0}^{\ast }]$ versus $%
\eta $).

\begin{figure}[htb]
  \centering
  \includegraphics[width=6.5cm]{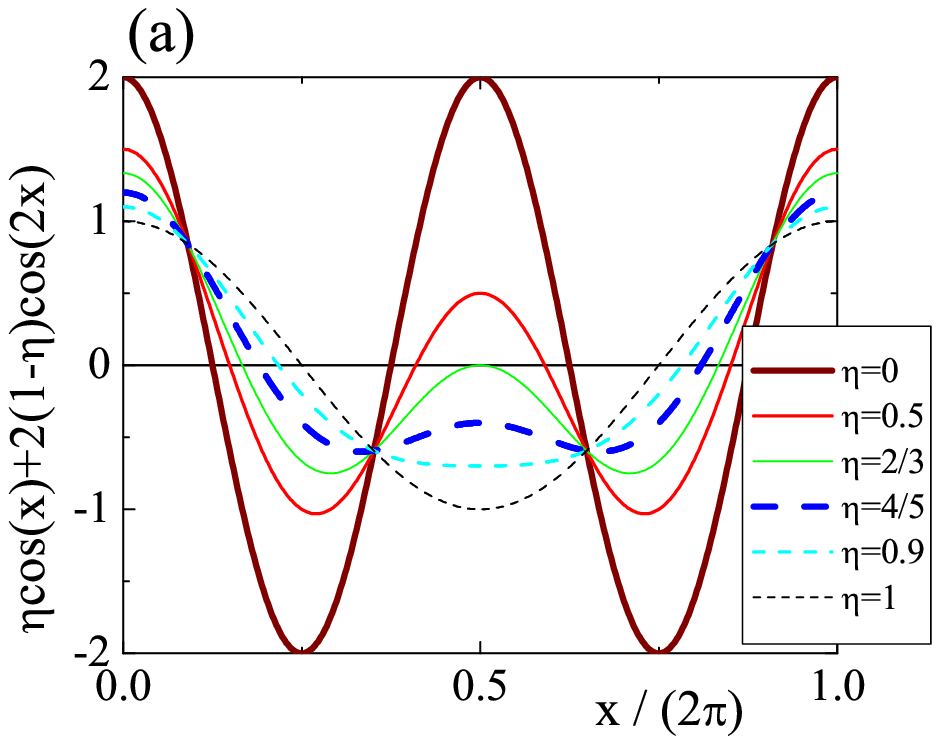}
  \includegraphics[width=6.5cm]{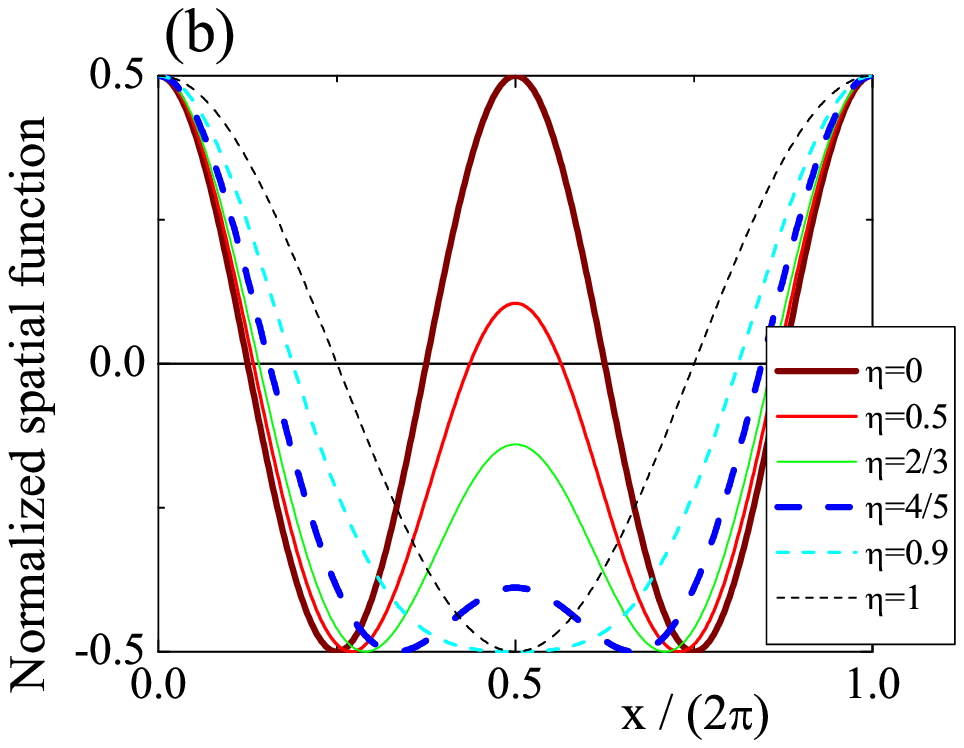}
  \caption{(Color online). (a) Spatial biharmonic function [Eq.~(2)] for the
 least favourable value $\varphi _{eff}=0$, and (b) the corresponding
 normalized function [Eqs.~(10) and (11)], versus space $x$ for different
 values of $\eta $.
  }
  \label{fig:three}
\end{figure}

\begin{figure}[htb]
  \centering
  \includegraphics[width=9.5cm]{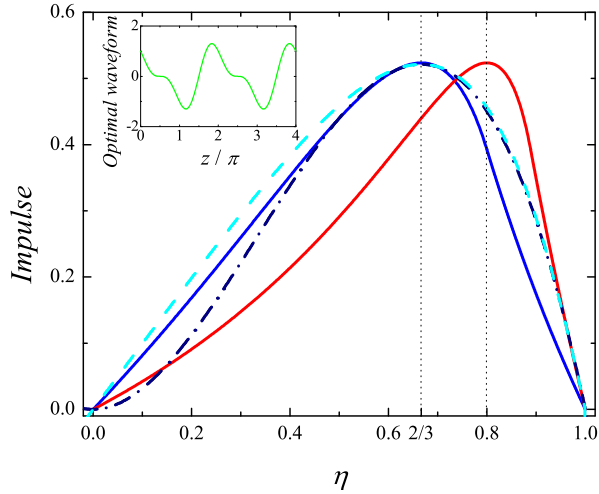}
  \caption{(Color online). Impulse functions $I[f_{\varphi _{eff}=0}^{\ast
  }]\left( \eta \right) $ [blue (black) line, Eq.~(9)],
  $I[g_{\varphi_{eff}=0}^{\ast }]\left( \eta \right) $ [red (gray) line, Eq.~(12)],
  $3.37(I[f_{\varphi _{eff}=\pi /2}^{\ast }]-1)$ [dashed line, cf.~Eq.~(6)],
  and $3.52\eta ^{2}\left( 1-\eta \right) $ (dashed-dotted line) versus $\eta $.
  The inset show the optimal waveform
  $\cos \left( z\right) -0.5\sin \left(2z\right) ,z=\left\{ t,x\right\} $,
  for both temporal and spatial biharmonic forces [Eqs.~(1) and (2)].
  }
  \label{fig:four}
\end{figure}

If the above scenario is universal (in a sense to be specified below), it
must also be found for the spatial force (2) since the forces (1) and (2)
exhibit the same symmetry properties, and hence the DSB is the same for both
types of forces. For the sake of comparison with the latter temporal case,
consider again the value $\varphi _{eff}=0$. In this case, one has to
renormalize the function $\eta \cos \left( x\right) +2\left( 1-\eta \right)
\cos \left( 2x\right) $ to change its image from $\left[ -\frac{33\eta
^{2}-64\eta +32}{16-6\eta },2-\eta \right] $ for $0\leqslant \eta \leqslant
8/9$ and $\left[ 2-3\eta ,2-\eta \right] $ for $8/9\leqslant \eta \leqslant
1 $ to $\left[ -1/2,1/2\right] $ for $0\leqslant \eta \leqslant 1$. The
resulting normalized function is now%
\begin{eqnarray}
g_{\varphi _{eff}=0}^{\ast }(x) &\equiv &\frac{g_{\varphi
_{eff}=0}(x)-\left( 2-\eta \right) /2-P\left( \eta \right) /2}{2-\eta
-P(\eta )},  \label{10} \\
P\left( \eta \right) &\equiv &\left\{
\begin{array}{c}
\frac{33\eta ^{2}-64\eta +32}{16\eta -16},\ 0\leqslant \eta \leqslant 8/9 \\
2-3\eta ,\ 8/9\leqslant \eta \leqslant 1%
\end{array}%
\right\} .  \label{11}
\end{eqnarray}%
Figure 3(b) shows a plot of $g_{\varphi _{eff}=0}^{\ast }(x)$. Accordingly,
the corresponding maximal transmitted impulse over a half-period is now%
\begin{equation}
I[g_{\varphi _{eff}=0}^{\ast }]\left( \eta \right) =\left( \frac{\pi }{2}%
\right) \frac{2-\eta +P\left( \eta \right) }{2-\eta -P\left( \eta \right) },
\label{12}
\end{equation}%
which exhibits a single maximum at $\eta =4/5$ with $I[g_{\varphi
_{eff}=0}^{\ast }]\left( \eta =4/5\right) =\pi /6$, i.e., the \textit{same}
value as $I[f_{\varphi _{eff}=0}^{\ast }]\left( \eta =2/3\right) $ (see
Fig.~4 for a plot of $I[g_{\varphi _{eff}=0}^{\ast }]$ versus $\eta $).
Observe that the generality of the present DRT scenario is additionally
supported by the remarkable conclusion that the two biharmonic forces (1)
and (2) have the \textit{same optimal waveform} at their respective optimal
values of $\eta $ and $\varphi _{eff}$ (see Fig.~4, inset). In this sense,
such an optimal waveform is universal, i.e., once one has identified the
relevant spatio-temporal symmetries to be broken in a given equation of
motion containing the force (1) or the force (2), this optimal waveform
maximally enhances the ratchet effect in that any other waveform (i.e., any
choice of the parameters $\eta $ and $\varphi _{eff}$ other than the
optimal) yields a lower ratchet effect while the remaining equation
parameters are held constant.

The present theory explains on the basis of a simple criticality scenario
all previously published results for a great diversity of systems (see, in
particular, the references cited in Ref. [14]). Additionally, this theory is
coherent with, and explains in a general setting, the general validity of
the scaling law $\left\langle V\right\rangle \sim \eta ^{2}\left( 1-\eta
\right) $ deduced in Refs. [14] and [17] for a temporal biharmonic force
with \textit{small} amplitudes. Indeed, one can now expect $\left\langle
V\right\rangle \sim S\left( \eta \right) $ for \textit{arbitrary}
amplitudes, where $S\left( \eta \right) $ exhibits features similar to those
of the functions $I[f_{\varphi _{eff}=0}^{\ast }]\left( \eta \right) $ and $%
I[g_{\varphi _{eff}=0}^{\ast }]\left( \eta \right) $ for temporal and
spatial biharmonic forces, respectively, i.e., $S(\eta =0,1)=0$ while $%
S\left( \eta \right) $ exhibits a single maximum at $\eta =2/3$ and $\eta
=4/5$ for temporal and spatial biharmonic forces, respectively. It also
explains: (i) the effectiveness of the traditionally used ratchet potential $%
V(x)=V_{0}\left[ \sin \left( 2\pi x/L\right) +0.25\sin \left( 4\pi
x/L\right) \right] $ [4-6]; (ii) the experimentally obtained optimal
parameters of the driving potential, biharmonic in both space and time, of a
quantum ratchet [18]; and (iii) the directed ratchet transport strength of
matter-wave solitons formed in a Bose-Einstein condensate [19]. Experimental
confirmation of the present findings can be readily obtained, for example in
the context of cold atoms in optical lattices [20]. A theory providing a
universal quantitative criterion to optimize the ratchet effect induced by
symmetry breaking of temporal and spatial biharmonic forces was
demonstrated, paving the way for any future optimal application of the
ratchet effect.

\begin{acknowledgments}
The author acknowledges useful conversations with M. Rietmann and R.
Carretero-Gonz\'{a}lez.
\end{acknowledgments}


\begin{thebibliography}{99}
\bibitem{15} Of course, this choice is also advantageous from an
experimental standpoint.
\end{thebibliography}
\end{document}